\documentclass[
    ,final            
  ,mathptmx                 
  ]
  {aipproc}

\layoutstyle{6x9}

\begin{document}

\title{A Complete Onium Program with \\ R2D at RHIC II}


\author{Richard Witt for the Exploratory Physics Working Group on a New Detector for RHIC II}
{
  address={Yale University, Wright Nuclear Structure Laboratory, 272 Whitney Ave., New Haven, CT 06520},
  email={Richard.Witt@yale.edu}
}

\begin{abstract}
Following on the discovery of a strongly interacting quark-gluon plasma (QGP)
at RHIC, a program of detailed quarkonia measurements is crucial to
understanding the nature of deconfinement.  Lattice QCD calculations suggest
a sequential melting of the quarkonia states in the deconfined medium.  Such
a melting would lead to a suppression in the measured charmonium and
bottomonium yields.  However, distinguishing a true suppression from shadowing,
absorption, and recombination effects requires detailed measurements of the charmonium 
states ($\mathrm{J}/\psi$, $\psi^{\prime}$, and $\chi_{\mathrm{c}}$) and bottomonium 
states ($\Upsilon$(1S), $\Upsilon$(2S), and $\Upsilon$(3S)).  Also, since measurements 
are needed not only in A+A, but also in $p$+$p$ for determining primary yields and in p+A 
for evaluating absorption, the detector should perform well in all collision environments.  
To fully realize the program outlined above, a new detector will be required at RHIC-II.  We 
present a proposal for a complete quarkonia program and the abilities of a new detector, R2D, 
to meet the stated requirements.  Comparisons will be made with proposed upgrades to existing 
RHIC detectors and with the upcoming LHC program.
\end{abstract}

\maketitle

\section{Why Onium?}

  Since the proposal of suppressed $\mathrm{J}/\psi$ production in heavy-ion collisions 
as a signature for the formation of a deconfined quark and gluon phase \cite{MatsuiSatz86}, 
ways have been sought to further exploit the probative value of heavy quark bound states.  
Lattice QCD calculations have shown an ordering of the binding energies of charmonium and 
bottomonium bound states from the $\mathrm{J}/\psi$ through the $\Upsilon$(3S) \cite{Karsch88}.  
This ordering suggests these states melt in a sequential manner that would allow a determination 
of the temperature of the deconfined state.  Table \ref{tab:Td} contains the binding energies ($E^{i}_{s}$) 
and disassociation temperatures ($T_{d}$), as a fraction of the critical temperature ($T_{c}$), for the various 
onium states.

\begin{table}
\begin{tabular}{l|rrr|rrrrr}
\hline
  \tablehead{1}{c}{b}{State}
  & \tablehead{1}{r}{b}{$\mathrm{J}/\psi$}
  & \tablehead{1}{r}{b}{$\chi_{\mathrm{c}}$}
  & \tablehead{1}{r}{b}{$\psi^{\prime}$}
  & \tablehead{1}{r}{b}{$\Upsilon$}
  & \tablehead{1}{r}{b}{$\chi_{\mathrm{b}}$}
  & \tablehead{1}{r}{b}{$\Upsilon^{\prime}$}
  & \tablehead{1}{r}{b}{$\chi_{\mathrm{b}}^{\prime}$}
  & \tablehead{1}{r}{b}{$\Upsilon^{\prime\prime}$} \\
\hline
$E^{i}_{s}$ [GeV] & 0.64 & 0.20 & 0.05 & 1.10 & 0.67 & 0.54 & 0.31 & 0.20 \\
\hline
$T_{d}/T_{c}$\tablenote{S. Digal, P. Petreczky, and H. Satz, Phys. Lett. B \textbf{514} (2001) 57} & 1.1 & 0.74 & 0.1--0.2 & 2.31 & 1.13 & 1.1 & 0.83 & 0.74 \\
$T_{d}/T_{c}$\tablenote{C.-Y. Wong, hep-ph/0408020} & $\sim 2.0$ & $\sim 1.1$ & $\sim 1.1$ & $\sim 4.5$ & $\sim 2.0$ & $\sim 2.0$ & -- & -- \\
\hline
\end{tabular}
\source{F. Karsch, http://rhicii-science.bnl.gov/heavy/doc/April05Meeting/karsch-lattice.pdf}
\caption{Binding energies, $E^{i}_{s}$, and disassociation temperatures, $T_{d}$, (as a fraction of the critical temperature $T_{c}$) 
for charmonium and bottomonium states from two lattice calculations.}
\label{tab:Td}
\end{table}

\section{Challenges}

  Results on $\mathrm{J}/\psi$ suppression in $\sqrt{s_{NN}}=200$ GeV were recently 
presented by the PHENIX collaboration \cite{PHENIXQM}.  In a sample of approximately $6.4\times10^{9}$ 
Au+Au events, PHENIX observes a factor of three suppression in the most central collisions relative to 
production in $p$+$p$ collisions at the same energy.  For comparison, NA50 reports a suppression factor of approximately 2 
with respect to Drell-Yan calculations in the most central Pb+Pb collisions at the SPS ($\sqrt{s_{NN}}=17.4$ GeV) 
\cite{NA50}.

However, an unambiguous interpretation of these results is complicated by several factors.  Firstly, the 
effects of nuclear shadowing and absorption must be taken into account.  The suppression factor of 2 measured 
by NA50 accounts for absorption by cold nuclear matter.  However, an alternative model that includes the effects of 
co-mover absorption is in very good agreement with the NA50 measurement \cite{Dinh02}.  Secondly, as pointed out by 
Grandchamp and Rapp, an understanding of the interplay of true QGP-related suppression and $c\overline{c}$ pair 
recombination is also important \cite{Grandchamp03}.  Consistancy checks are needed to ensure that the charm and bottom 
quarks freed by onium disassociation are found elsewhere in the final state (flavor conservation).  Lastly, the 
$\chi_{\mathrm{c}}$ has a substantial $\mathrm{J}/\psi$ radiative decay channel.  A simple projection of the NRQCD 
calculations that fit the recent HERA-B $\chi_{\mathrm{c}}$ measurement suggests as much as 70\% of the $\chi_{\mathrm{c}}$ 
decay radiatively to $\mathrm{J}/\psi$ at RHIC energies \cite{HERAB}.  The $\chi_{\mathrm{c}}$ has a much lower $T_{d}$ than the 
$\mathrm{J}/\psi$, hence melting of the $\chi_{\mathrm{c}}$, which occurs below $T_{c}$, will result in an artificially 
suppressed $\mathrm{J}/\psi$ yield.  Measurements of the $\chi_{\mathrm{c}}$ production at RHIC energies are required to 
fully understand the contribution of the $\chi_{\mathrm{c}}$ feeddown to the observed $\mathrm{J}/\psi$ suppression.

\section{Meeting the Requirements}

A comprehensive new detector will be necessary to perform all the measurements required for a complete onium 
program at RHIC II.  The basic requirements can be summarized as follows.  The baseline measurements, \textit{i.e.} 
all charmonium and bottomonium states mentioned above in A+A, $p$+A, and $p$+$p$ collisions as a function of centrality, 
transverse momentum, rapidity, $\sqrt{s_{NN}}$, $x_{F}$, and beam species, will require very high collision rates and 
a large acceptance to take sufficient statistics in the available beam time.  High mass resolution (approximately 60 MeV/$c^2$) 
is required to resolve the three $\Upsilon$ states.  Large acceptance and forward rapidity coverage is needed to reach 
large $x_{F}$ and low $x_{BJ}$.  A high resolution vertex determination and full, symmetric azimuthal coverage is required 
for correlation and elliptic flow measurements that will allow us to disentangle recombination effects and actual 
suppression.  Lastly, the $\chi_{\mathrm{c}}$ measurement cannot be done without photon detectors at both mid and forward 
rapidities.

We have proposed a new detector, R2D, that meets or surpasses all of the above requirements \cite{R2D}.  Two possible 
baselines have been explored, one based on the SLD magnet, the other on the CDF magnet.  In either configuration R2D will 
offer high precision tracking, PID, and EM and hadronic calorimetery all over $|\eta|<3.5$ and $\Delta\phi=2\pi$.  
The mass resolution will be approximately 60 MeV/$c^2$, sufficient 
to resolve the $\mathrm{J}/\psi$, the $\psi(2\mathrm{S})$, and all three $\Upsilon$ states in both dilepton channels.  By 
way of comparison, PHENIX will only be able to achieve a 60 MeV/$c^2$ mass resolution in the electron channel after its VTX 
upgrade.  STAR will have a 170 MeV/$c^2$ mass resolution in the electron channel even after a possible $\mu$-vertex (HFT) upgrade.  
R2D will also provide a forward ($3.5<|\eta|<4.8$) spectrometer section with silicon tracking, particle ID, hadronic and 
electromagnetic calorimetry, and a muon chamber.  A more recent alternative configuration would provide the necessary 
detector capabilities to handle electron-ion collisions in the \textit{e}RHIC era.  A comparison of the expected rates from 
PHENIX, STAR, ALICE, CMS, and R2D is presented in Table \ref{tab:Frawley}.

\begin{table}
\begin{tabular}{l|r|rrrrr}
\hline
  \tablehead{1}{c}{b}{Signal}
  & 
  & \tablehead{1}{r}{b}{PHENIX}
  & \tablehead{1}{r}{b}{STAR}
  & \tablehead{1}{r}{b}{ALICE}
  & \tablehead{1}{r}{b}{CMS}
  & \tablehead{1}{r}{b}{R2D} \\
\hline
$\mathrm{J}/\psi \rightarrow \mu^{+}\mu^{-}$ or $e^{+}e^{-}$ & $p$+$p$ & 525,000 & 1,600,000 & 135,900 & 17,219 & 10,200,000 \\ 
& A+A & 440,000 & 220,000 & 749,500 & 24,000 & 8,580,000  \\
\hline
$\psi \rightarrow \mu^{+}\mu^{-}$ or $e^{+}e^{-}$  & $p$+$p$ & 9,490 & 29,000 & 2,450 & 310 & 184,000 \\ 
& A+A & 7,900 & 4,000 & 14,190 & 440 & 154,000  \\
\hline
$\chi_{\mathrm{c}} \rightarrow \mu^{+}\mu^{-}\gamma$ or $e^{+}e^{-}\gamma$  & $p$+$p$ & 142,600 & -- & -- & -- & 1,560,000 \\
& A+A & 119,900 & -- & -- & -- & 1,320,000 \\
\hline
$\Upsilon$ (unresolved)  & $p$+$p$ & 700 & 8,300 & 1,350 & 3,010 & 35,200 \\
& A+A & 1,440 & 11,200 & 11,000 & 26,000 & 71,000 \\
\hline
$\Upsilon$ (resolved) & $p$+$p$ & 210 & 0 & 1,350 & 3,010 & 35,200 \\
& A+A & 400 & 0 & 11,000 & 26,000 & 71,000 \\
\hline
$\mathrm{B} \rightarrow \mathrm{J}/\psi \rightarrow \mu^{+}\mu^{-}$ or $e^{+}e^{-}$  & $p$+$p$ & 3,300 & 19,000 & 3,580 & 573 & 68,000 \\
& A+A & 6,270 & 2,500 & 12,900 & 2,060 & 132,000 \\
\hline
\end{tabular}
\source{A. Frawley, http://rhicii-science.bnl.gov/heavy/doc/white-paper/wp-draft-jan12.pdf}
\caption{Comparison of expected rates at RHIC-II for charmonium and bottomonium states in $\mu^{+}\mu^{-}$ and $e^{+}e^{-}$ 
channels from the current RHIC detectors (including planned upgrades), the LHC detectors, and R2D.  Entries with ``--'' indicate that the 
information available does not allow reliable estimation to be made.  Estimates are for 12 weeks for RHIC-II $p$+$p$ (238 pb$^{-1}$)
and A+A (18 nb$^{-1}$) at $\sqrt{s_{NN}}=200$ GeV, and 1 month for LHC $p$+$p$ (3 pb$^{-1}$) and A+A (500 $\mu$b$^{-1}$) 
at $\sqrt{s_{NN}}=5.5$ TeV.}
\label{tab:Frawley}
\end{table}


\end{document}